\documentclass[12pt]{article}
\makeatletter

\def\hatgap{2pt}
\def\subdown{-2pt}
\newcommand\reallywidehat[2][]{%
\renewcommand\stackalignment{l}%
\stackon[\hatgap]{#2}{%
\stretchto{%
    \scalerel*[\widthof{$#2$}]{\kern-.6pt\bigwedge\kern-.6pt}%
    {\rule[-\textheight/2]{1ex}{\textheight}}
}{0.5ex}
_{\smash{\belowbaseline[\subdown]{\scriptstyle#1}}}%
}}

\usepackage{subfigure}

\mathchardef\ordinarycolon\mathcode`\:
\mathcode`\:=\string"8000
\begingroup \catcode`\:=\active
  \gdef:{\mathrel{\mathop\ordinarycolon}}
\endgroup

\usepackage{rotating}

\usepackage{setspace}
\usepackage{times}

\usepackage{bm}
\usepackage{epsfig}
\usepackage{graphicx}
\usepackage{epstopdf}
\usepackage{amsmath}
\usepackage{multirow}

\usepackage{amssymb}
\usepackage{tikz}
\usetikzlibrary{positioning}

\newcounter{lastnote}

\title{Electoral Stability and Rigidity}

\author{Michael Y. Levy$^{1,2\ast}$\\
\normalsize{$^{1}$New York City College of Technology} \\
\normalsize{186 Jay Street,  V633} \\
\normalsize{Brooklyn, New York 11201- 2983} \\
\normalsize{}\\
\normalsize{$^{2}$Honeycomb Money Management}\\
\normalsize{Brooklyn, New York 11238}\\
}

\date{}

\begin{document} 
\maketitle 
\begin{abstract}
\baselineskip11pt
Some argue that political stability is best served through a two-party system. This study refutes this. The author mathematically
defines the stability and rigidity of electoral systems comprised of any
quantity of electors and parties. In fact, stability is a function of the quantity of electors - i.e., the number of occupied seats at the table. As the number of electors increases, the properties of an electorate are increasingly well resolved, and well described by those of an electorate that is least excessive -- that is to say an electorate that is closest to equilibrium. Further, electoral rigidity is a function of the quantity of parties and their probabilities of representation. An absolutely rigid system admits no fluctuations -- whatever happens to one elector will happen to all electors. As the quantity of parties increases so does the number of party lines, and with it the quantity of alternatives with which to respond to an external stimulus. Rigidity is significant in a social system that places high value on party loyalty. In conclusion,  (\emph{i}) electoral stability is best served by increasing the quantity of electors;  (\emph{ii}) electoral rigidity is best served by decreasing the quantity of parties, and by increasing the representation of some parties at the expense of others; and  (\emph{iii}) the less stable a branch of government, the more concern is placed on those who would hold those offices for the people.
\end{abstract}
\section*{Acknowledgments}
This manuscript is dedicated to the public. The author acknowledges David Citrin for critically reading the manuscript; 
Esther Levy and Phoebe Levy for editing the manuscript; Sean Jacobs and Anonymous for supporting the data acquisition; and the University of New Mexico, the Santa Fe Institute, the New York University, and the New York Public Library's Manhattan Research Library Initiative for providing library access. 

\clearpage

\baselineskip24pt

\section{Introduction}
In Timmons versus Twin-Cities-Area New Party, the U.S. Supreme Court considered the constitutionality of the ``many barriers in the political arena\cite{renquist-1997}.'' The Twin-Cities-Area New Party argued that its associational rights were violated by Minnesota electoral law. Writing for the majority, Renquist thinks ``the Constitution permits the Minnesota Legislature to decide that political stability is best served through a healthy two-party system\cite{renquist-1997}\footnote{The court's majority opinion implies that the stability, $\mathcal{S}$, is a function, $f$, of the quantity of parties, $M$ (i.e. $\mathcal{S} = f{\left(M\right)}$).  This manuscript clarifies that the stability is a function, $g$, only of the quantity of electors, $N$ (i.e. $\mathcal{S} = g{\left(N\right)} \neq f{\left(M\right)}$); and that that the rigidity, $\mathcal{R}$, is a function, $h$, of the quantity of parties, $M$ (i.e. $\mathcal{R} = h{\left(M\right)}$).}.'' Though what the U.S. Constitution permits or does not permit with regard to associational rights is outside the present scope, the author draws on techniques well known in statistical mechanics to refute the thought that premises Timmons. The author concludes that electoral stability and rigidity are best served by increasing the quantity of electors and decreasing the quantity of parties, respectively. 

Electoral science has a long tradition of combining ethics and reason to study how choice is resolved. With respect to ethics, Black\cite{black-1958} recounts two conundra such that in a close election\footnote{The electoral theory herein  holds for any election or selection so long as the vote or choice may be given as an integer partion of $N$ into $M$ parts (see Section~\ref{sec:vote-configurations}).  Thus, the formalism here applies to most any vote by any electoral body regardless of who the electors are (i.e., citizens, committees, juries, executives, judges, legislators, etc.) or the subject matter(s) under consideration.  Four relevant applications of this theory are discussed in Section~\ref{sec:discussion}.} between people, motions, or parties, the election may be with an unjust resolution or without any resolution. Specifically, Borda\cite{borda-1781} describes that the party with the highest number of votes may not be the most preferred party and Condorcet\cite{condorcet-1785} describes a three-way draw. Though several authors\cite{borda-1781,condorcet-1785, laplace-1812,dodgson-1873,dodgson-1874,dodgson-1876,nanson-1882,dodgson-1884,hoag-1926} prescribe rules to reduce these likelihoods, Arrow\cite{arrow-1951} proves that it is impossible to preclude them. Ethics aside, Key\cite{key-1955} and Campbell\cite{campbell-1966} explain a theory of critical elections whereby the electorate's direction is flipped or rotated as compared to antecedent elections, and moreover the re-orientation persists for one or more subsequent elections. These latter two studies are of particular relevance to the present study because they employ directionality in a qualitative manner to describe the electorate, whereas herein the author uses vectors, which include both magnitude and direction, to study the electorate in a quantitative manner. 

The problem addressed here is to understand how the stability and rigidity of an electoral system vary with the quantity of electors, $N$, and parties, $M$. The philosophical approach (Section~\ref{sec:phil-approach}) is to adhere to the law of parsimony by utilizing an isotropic preference structure. The technical approach (Section~\ref{sec:approach}) is to understand multi-party electoral systems by utilizing multinomial properties and both magnitudal and directional ensemble averages. The results (Section~\ref{sec:results}) are explained qualitatively by using two visual aids (Figures~\ref{fig:results-stability} and \ref{fig:results-rigidity}), and quantitatively by writing closed-form analytic expressions for the stability (Equation~\ref{eq:stability-closed-form}) and rigidity (Equation~\ref{eq:rigidity-closed-form}) of electoral systems. Results indicate that (i) beginning with absolute instability for a single-elector system, the stability increases monotonically to unity as the quantity of electors, $N$, increases; and (ii) beginning with absolute rigidity for a one-party system, the rigidity decreases monotonically to null as the quantity of parties, $M$, increases. The discussion (Section~\ref{sec:discussion}) focuses on asking and answering the following question: how does this theory of electoral dynamics enhance the understanding of a real electoral systems. This discussion makes use of the constitutional order of the United States of America as an archetype. Each branch of government is analyzed individually and then they are analyzed together comparatively. Finally, the author draws three conclusions (Section~\ref{sec:conclusions}): (1) electoral stability is best served by increasing the quantity of electors; (2) electoral rigidity is best served by decreasing the quantity of parties, and decreasing their proportional representation; and (3) the less stable a branch of government, the more concern is placed on those who would hold those offices for the people.  There are two appendices.  Each provides a necessary mathematical proof.

\section{Philosophical Approach\label{sec:phil-approach}}
The author's approach is grounded in the canon of philosophical thought. Throughout this study of electoral dynamics the author adheres to Ockham's law of parsimony: ``explain[ing] by the assumption of fewer things~\cite{ockham-1300},'' while adequately representing the most salient properties. As a consequence, the author utilizes an isotropic preference structure -- the underlying preference is uniformly likely to be in any direction -- and measures the properties of the entire ensemble of permissible preference structures. A description of any one structure is foregone in favor of a gross description of all structures simultaneously. 

This approach is idealistic and realistic. Conventional wisdom dictates that ``politics makes strange bedfellows\cite{warner-1880}.'' Thus, no matter how seemingly unlikely an electoral alliance may appear, even the least preferable alliance may not be ignored or discounted. Further, we empirically know that there are numerous ways for Party~A and Party~B to be oriented. At any given time, Party~A and Party~B may be in agreement (parallel orientation) and be either in favor of or against electoral issue $i$; they may be in disagreement and be in opposition (anti-parallel orientation) on electoral issue $j$; either party may be in favor of or against electoral issue $k$ while the other party is ambivalent\cite{farrier2010congressional,craig2005ambivalence} on the issue (perpendicular orientation); or, they may both be ambivalent to electoral issue $l$, but for different reasons (co-planar orientation). 

\section{Technical Approach\label{sec:approach}}
The technical approach is to expand from an understanding of a binary system\cite{kittel-1980}, whose states are described with a binomial expansion, to an understanding of a multi-party system, whose states are described by a multinomial expansion. The author begins by specifying the model that is used to represent electoral systems (Section~\ref{sec:specification-model}). The author continues by writing the generating function, which yields the probability of each and every vote by an electorate (Section~\ref{sec:generating-function}). Then, the author defines the primary vectors that will be used throughout this study; chief among them are the electoral vector and the equilibrium-electoral vector (Section~\ref{sec:vector-definitions}). The author then briefly uses number theory to explain that each and every vote taken by an electorate is a permutation of the partitions of a positive integer $N$ into $M$ (Section~\ref{sec:vote-configurations}). The author explains the relationship between the multiplicity of each vote configuration and the probability-density function of a multinomially distributed variable (Section~\ref{sec:multiplicity-function}). Next, the author explains that a quantitative understanding of an electoral dynamics necessitates the use of two ensemble averages: a magnitudal and a directional average (Section~\ref{sec:ensemble-averages}). The author then uses vector algebra to define the fractional fluctuation, which a relative measure of the expected deviation of a system from its equilibrium (Section~\ref{sec:fractional-fluctuation}). The author explains that the fractional fluctuation is separable, and may be written as the product of the volatility, which is solely a function of the quantity of electors, and the flexibility, which is solely a function of the quantity of parties; and that the stability and rigidity are the unitary complements of the volatility and flexibility, respectively (Section~\ref{sec-four-dynamic-variables}). Two assist the reader in comprehending this material, the author provides an example electoral system in Figure~\ref{fig:model_system}, several examples of integer partitions of $N$ into $M$ in Table~\ref{tab:one--}, and a thorough example of how to calculate an ensemble average in Table~\ref{tab:two-}.

\subsection{\label{sec:specification-model}Specification of the Model}
The author enumerates this theory with eleven statements, all of which are given below
\begin{enumerate}
\renewcommand{\theenumi}{\alph{enumi}}
\item 
There is a positive integer number, $N$, of seat of power in the system.
\item 
An elector occupies each and every seats of power.
\item 
There is a fixed integer $M$ of possible outcomes. 
\item 
The direction of the $k^\textrm{th}$ possible outcome points in the direction of unit vector $\boldsymbol{\hat{k}}$.
\item
Each elector must align with one and only one outcome.
\item 
The event probabilities of the outcomes are $p_1, p_2, \ldots, p_M$,
\item 
An elector aligned with the $k^\textrm{th}$ outcome will have an electoral moment $\boldsymbol{\bar{m}}_k$. 
\item 
Each permutation of the electoral marks is electorally allowed.
\item 
Each orientation of the electoral moments is electorally allowed.
\item 
The directional probabilities of orientation are uniformly distributed over all orientations.
\item 
The direction of the electoral moment $\boldsymbol{\bar{m}}_k$ is isotropic for $k = 1,2, \ldots, M$.
\end{enumerate}

\subsubsection{\label{sec:example-electoral-system}Example Electoral System}
\begin{figure}[hbt!]
\begin{center}
\includegraphics[width=10.0cm]{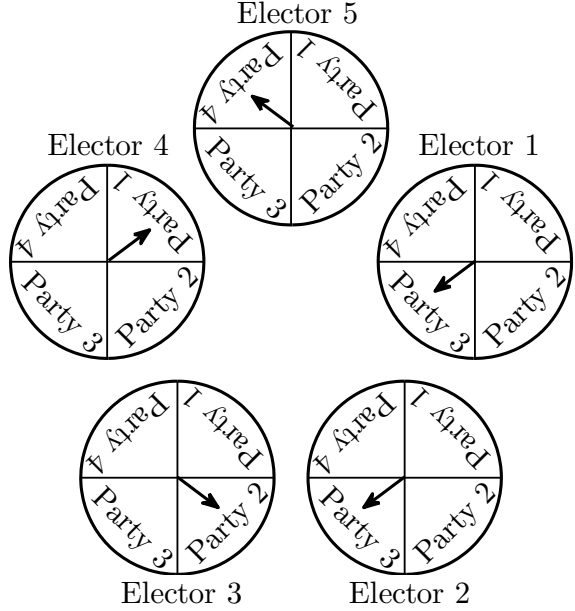}
\caption{\label{fig:model_system}Example electoral system. Here the electoral system is comprised of five ($N=5$) electors whose moments, $\boldsymbol{\overline{m}_i}$, points in one of four ($M=4$) directions. The state illustrated here has two electors pointing in the direction of the Party 3, whereas there is one elector pointing in the direction of the remaining parties; thus, the electoral vector of this state is $\boldsymbol{\overline{N}} = \left(1\right) \, \boldsymbol{\hat{1}} + \left(1\right)\, \,\boldsymbol{\hat{2}} + \left(2\right) \,\boldsymbol{\hat{3}} + \left(1\right) \, \boldsymbol{\hat{4}}$.   Though it may appear in this figure that $\boldsymbol{\hat{3}} = - \boldsymbol{\hat{1}}$, in reality these vectors exist in a multidimensional vector space.  The interested reader may find more information in Appendix~\ref{sec:Proof-A}.}
\end{center}
\end{figure}

An example of the model electoral system is illustrated in Figure~\ref{fig:model_system}. There the reader finds five seats of power, each occupied with an elector whose single arrow points is aligned with one of four distinct parties. The state of the example system is such that there is one elector aligned with party 1, one elector aligned with party 2, two electors aligned with party three, and one elector aligned with party 4.

\subsection{\label{sec:generating-function}Generating Function}
Each and every permutation of a vote by an electorate is generated by the symbolic product of $N$ factors as
\begin{equation}
\left( p_1\,\boldsymbol{\hat{1}} + p_2\,\boldsymbol{\hat{2}} + \ldots + p_M\,\boldsymbol{\hat{M}}\right)^N \textrm{.}
\end{equation}
Each factor represents an elector that is directed in exactly one of some fixed finite number $M$ possible directions, $\boldsymbol{\hat{1}}, \boldsymbol{\hat{2}}, \ldots \boldsymbol{\hat{M}}$, with probabilities $p_1, p_2, \dots , p_M$. Consistent with electoral reality, these probabilities are bounded as $0 \leq p_i \leq 1$ and constrained as 
\begin{equation}
\label{eq:contraint-equation}
\sum_{i=1}^{M}{p_i} = 1.
\end{equation}
 On multiplication, there are $M^N$ terms, one each for the $M^N$ permutations of the marks. The set of directional unit vectors $\boldsymbol{\hat{1}}, \boldsymbol{\hat{2}}, \ldots \boldsymbol{\hat{M}}$ need not form an orthogonal set (see Appendix~\ref{sec:Proof-A}).

\subsection{\label{sec:vector-definitions}Vector Definitions}
Each state of the system has a correspondence with an electoral-excess vector, $\boldsymbol{\overline{N}}$, that describes it as
\begin{equation}
\label{eq:electoral-vector}
\boldsymbol{\overline{N}} := \sum\limits_{k=1}^M{ N_k \,\boldsymbol{\hat{k}}} \, \textrm{,}
\end{equation} 
where $N_k$ is the number of electors aligned with the $k^{\textrm{th}}$ party.
Additionally, each system has an equilibrium-electoral vector that is given as
\begin{equation}
\label{eq:equilibrium-vector}
\boldsymbol{\overline{N}_o} := N\,\sum\limits_{k=1}^M{p_k\,\boldsymbol{\hat{k}}}\,.
\end{equation}
From these two, the author defines the electoral-excess vector as
\begin{equation}
\label{eq:electoral-excess-vector}
\boldsymbol{\overline{x}}:= \boldsymbol{\overline{N}} - \boldsymbol{\overline{N}_o} = \sum\limits_{k=1}^M{ \left(N_k - p_k\,N\right) \,\boldsymbol{\hat{k}}} \, \textrm{,}
\end{equation} 
The last vector to be defined is the one vector, $\boldsymbol{\bar{1}}$. It is given as a linear combination of the directions of all the outcomes as
\begin{equation}
\boldsymbol{\overline{1}} := \sum\limits_{k=1}^M{\boldsymbol{\hat{k}}}\,.
\end{equation}
The reader recalls that $\boldsymbol{\hat{k}}$ is the orientation of the $k^{\textrm{th}}$ party, and $p_k$ is the event probability of outcome $k$. 

\subsection{\label{sec:vote-configurations}Vote Configurations}
Irrespective of each elector's underlying preference, the result of a vote is always a partition~\cite{leibniz, euler} of $N$ into exactly $M$ parts\footnote{The theory here can be exteneded to allow for abstention or non-participation.  For example, allow that from the $N$ seats of power that $N_A$ electors abstain.  Including for this eventuality, the system has one additional outcome.  Thus by taking into account absentions, a vote is always a partition of $N$ into exactly $M+1$ parts.}. 
In Table~\ref{tab:one--}, the author presents five complete set of partitions; one each for five distinct electoral systems. In all cases, $N = 5$, and $M = 1, 2, \ldots 5$, respectively. 
The cardinality, $c{\left(N,M\right)}$, of each set, $C{\left(N,M\right)}$, is written on the last row of the table.
\begin{table}[ht]
\begin{center}
\caption{\label{tab:one--}
Example of partitions of $N$ into $M$. Irrespective of each elector's preferential marks, the vote must tally to a partition of $N$ into $M$. The cardinality, $c{\left(N,M\right)}$, of each set of partitions, $C{\left(N,M\right)}$, is given on the bottom-most row.}
\renewcommand*{\arraystretch}{1.4}
\vspace{0.3em}
\begin{tabular}{|c|c|c|c|c|c|c|}
\hline
$N = 5, M = 1$ & $N = 5, M = 2$ & $N = 5, M = 3$ & $N = 5, M = 4$ & $N = 5, M = 5$ \\
\hline 
$\left[5\right]$ & $\left[5, 0\right]$ & $\left[5, 0, 0\right]$ & $\left[5, 0, 0, 0\right]$ & $\left[5, 0, 0, 0, 0\right]$ \\
\hline 
& $\left[4, 1\right]$ & $\left[4, 1, 0\right]$ & $\left[4, 1, 0, 0\right]$ & $\left[4, 1, 0, 0, 0 \right]$ \\
\hline 
& $\left[3, 2\right]$ & $\left[3, 2, 0\right]$ & $\left[3, 2, 0,0\right]$ & $\left[3, 2, 0,0, 0\right]$ \\
\hline 
& & $\left[3, 1, 1\right]$ & $\left[3, 1, 1, 0\right]$ & $\left[3, 1, 1, 0, 0\right]$ \\
\hline 
& & $\left[2, 2, 1\right]$ & $\left[2, 2, 1, 0\right]$ & $\left[2,2, 1, 0, 0\right]$ \\
\hline 
& & & $\left[2, 1, 1, 1\right]$ & $\left[2, 1, 1, 1, 0\right]$ \\
\hline 
& & & & $\left[1, 1, 1, 1, 1\right]$ \\
\hline 
\hline 
$c{\left(N,M\right)}$
= 1 & $c{\left(N,M\right)}$
= 3 & $c{\left(N,M\right)}$
= 5 & $c{\left(N,M\right)}$
= 6 & $c{\left(N,M\right)}$
= 7 \\
\hline 
\end{tabular}
\end{center}
\end{table}

\subsection{\label{sec:multiplicity-function}Multiplicity Function}
Multiplicity functions are central to this study. The multiplicity function, $g{\left(\boldsymbol{\overline{N}} \right)}$ gives the number of occurrences of the electoral vector $\boldsymbol{\overline{N}}$. Within a constant of proportionality, the multiplicities are given by the probability-mass function of the multinomial distribution as 
\begin{equation} 
\label{eq:multiplicity}
g{\left(\boldsymbol{\overline{N}} \right)} = M^N\, N! \prod\limits_{k=1}^M{{\frac{ p_k^{N_k}}{N_k!}}} \, \textrm{.} 
\end{equation}
Through graphic presentations of the multiplicity functions (see Figures~\ref{fig:results-stability} and \ref{fig:results-rigidity}), the reader may respectively obtain a complete qualitative and quantitative understanding of how the stability and rigidity of electoral systems vary with respect to the quantity of electors and parties, respectively. 

In Table~\ref{tab:two-} the author includes relevant information regarding an electoral system with $N=5$, $M=3$, $p_1 = 0.1$, $p_2 = 0.3$, and $p_3 = 0.6$. In column 1, the author writes each permutation of each of the $c{\left(N=5,M=3\right)}$ partitions in the set $C{\left(N=5,M=3\right)}$. The number of permutations of the vector $\boldsymbol{\bar{N}}$ is $d{\left(\boldsymbol{\bar{N}}\right)}$ where
\begin{equation}
d{\left(\boldsymbol{\bar{N}}\right)}
= \frac{M!}{\prod_{n=0}^N{m{\left(n\right)!}}}\,\textrm{.}
\end{equation}
In the above equation, $m{\left(n\right)}$ is the multiplicity of integer $n$ -- i.e., the number of instances of $n$ in the multiset associated with vector $\boldsymbol{\bar{N}}$. In column 2, the author writes the multiplicity, $g{\left(\boldsymbol{\overline{N}}\right)}$, of each permutation. The sum of multiplicities, $g{\left(\boldsymbol{\overline{N}}\right)}$, correctly totals to $M^N$. 
\begin{table}[hp!]
\begin{center}
\caption{\label{tab:two-}
Example: $N=5, M=3, p_1 = 0.1, p_2 = 0.3, p_3 = 0.6$. To calculate the expected value of a dynamic variable, each and every electoral state must be included (column 1). For each state, the multiplicity (column 2) and the dynamic variable (column 3) are calculated From these, a weighted sum is computed (column 4). Finally, the weighted sum is divided by the total number of states. For an electoral system with five electors ($N = 5$) and three parties ($M = 3$), the expected value of the systems variance is 656.1/ 243 }
\renewcommand*{\arraystretch}{1.35}
\vspace{0.3em}
\begin{tabular}{|c|c|c|c|c|c|c|}
\hline
$\boldsymbol{\overline{N}}$ & $g{\left(\boldsymbol{\overline{N}}\right)}$& $\left\{ \left(\boldsymbol{\overline{N}} - \boldsymbol{\overline{N}}_o \right) \boldsymbol{\cdot} \left(\boldsymbol{\overline{N}} - \boldsymbol{\overline{N}}_o \right) \right\} $ & $g{\left(\boldsymbol{\overline{N}}\right)}\,\left\{ \left(\boldsymbol{\overline{N}} - \boldsymbol{\overline{N}}_o \right) \boldsymbol{\cdot} \left(\boldsymbol{\overline{N}} - \boldsymbol{\overline{N}}_o \right) \right\} $ \\
\hline 
$\left[5, 0, 0\right]$ & 0.0 & 32 & 0.1 \\
\hline 
$\left[0, 5, 0\right]$ & 0.6 & 22 & 12.7 \\
\hline 
$\left[0, 0, 5\right]$ & 18.9 & 7 & 122.8 \\
\hline 
$\left[4, 1, 0\right]$ & 0.0 & 22 & 0.8 \\
\hline 
$\left[4, 0, 1\right]$ & 0.1 & 19 & 1.4 \\
\hline 
$\left[1, 4, 0\right]$ & 1.0 & 16 & 15.3 \\
\hline 
$\left[1, 0, 4\right]$ & 15.7 & 4 & 55.1 \\
\hline 
$\left[0, 4, 1\right]$ & 5.9 & 11 & 62.0 \\
\hline 
$\left[0, 1, 4\right]$ & 47.2 & 2 & 70.9 \\
\hline 
$\left[3, 2, 0\right]$ & 0.2 & 16 & 3.4 \\
\hline 
$\left[3, 0, 2\right]$ & 0.9 & 10 & 8.3 \\
\hline 
$\left[ 2, 3, 0\right]$ & 0.7 & 14 & 8.9 \\
\hline 
$\left[2,0,3\right]$ & 5.2 & 5 & 23.6 \\
\hline 
$\left[0, 3, 2\right]$ & 23.6 & 4 & 82.7 \\
\hline 
$\left[0, 2, 3\right]$ & 47.2 & 0 & 23.6 \\
\hline 
$\left[3, 1, 1\right]$ & 0.9 & 11 & 9.2 \\
\hline 
$\left[1, 3, 1\right]$ & 7.9 & 7 & 51.2 \\
\hline 
$\left[1, 1, 3\right]$ & 31.5 & 1 & 15.8 \\
\hline 
$\left[2, 2, 1\right]$ & 3.9 & 7 & 25.6 \\
\hline 
$\left[2, 1, 2\right]$ & 7.9 & 4 & 27.6 \\
\hline 
$\left[1, 2, 2\right]$ & 23.6 & 2 & 35.4 \\
\hline 
\hline 
Sum & $243\,(3^5)$ & & 656.1 \\
\hline 
\end{tabular}
\end{center}
\end{table}

Before continuing to the next section the reader will please note that each variable $N_k$ ($k = 1,2\ldots,M$; see Equation~\ref{eq:electoral-vector}) possesses a binomial distribution. For example, with the aid of Table~\ref{tab:two-} the reader can verify that the sum of all multiplicities, $g{\left(\boldsymbol{\overline{N}}\right)}$, for each tuple where $N_1 = 1$ is 79.7 (i.e., $1.0 + 15.7 + 7.9 + 31.5 + 23.6 = 79.7$). This sum divided by $M^N$ is 0.33 (i.e., $79.7/3^5 = 0.33$). This quotient is equal to the probability-density function of a binomially distribution variable with event probability $p_1 = 0.1$ that is evaluated at $N_1 = 1$. Note that the probability-density function, $P$, of a binomially distributed variable, $N_k$, evaluated at an integer value $n$ is 
\begin{equation}
P{\left(N_k = n\right)} = \frac{N!}{n!\,\left(N-n\right)!}\, p_k^n\,\left(1- p_k\right)^{N-n}, \textrm{ for } n = 0,1, \ldots, N.
\end{equation}
Thus, the author utilizes the binomial distribution to qualitatively describe the stability and rigidity of electoral systems (see the comments in Section~\ref{sec:qualitative-results} regarding Figures~\ref{fig:results-stability} and \ref{fig:results-rigidity}).

\subsection{\label{sec:ensemble-averages}Ensemble Averages}
A complete quantitative understanding of the stability and rigidity of electoral systems requires the use of ensemble averages. In this paper, the author utilizes two distinct ensemble averages of a function $f$: a magnitudal average, $\left<f\right>$, and a directional average, $\left\{f\right\}$. 

\subsubsection{Directional Ensemble Average} 
The directional ensemble average is given as 
\begin{equation}
\label{eq:Directional-Ensemble-Average}
\left\{ f \right\} 
= 
\frac{\int_{\phi_{M-1}}^{2\,\pi}\,\int_{\phi_{M-2}}^{\pi}\, \cdots \, \int_{\phi_1}^{\pi} f \,\, d^{M-1}S 
}{\int_{\phi_{M-1}}^{2\,\pi}\,\int_{\phi_{M-2}}^{\pi}\, \cdots \, \int_{\phi_1}^{\pi} d^{M-1}S 
} \textrm{,}
\end{equation}
where $d^{M-1}S$ is the differential surface area of an $M$-sphere and $\phi_1, \phi_2, \ldots, \phi_{M-1}$ are the angular coordinates of the $M$-dimensional\footnote{Most generally speaking, there is one-dimension for each of the fixed integer $M$ possible outcomes.  In the special case of party dynamics, the fact that two or more parties may agree on one issue does not lower the dimensionality of the system in the broader context.  De facto, each party has its own unique genesis and thus, in broad terms, augments the dimensionality of an electoral system.} electoral vector space (see Appendix~\ref{sec:Proof-A} for more information). As an example of use, in column 3 of Table~\ref{tab:two-} the author provides a directional ensemble average for each permutation of each partition. Specifically, the author gives the directional ensemble average of the dot product of the excess vector with itself, where the electoral-excess vector is equal to the difference between the electoral vector and the equilibrium-electoral vector.

\subsubsection{Magnitudal Ensemble Average}
The magnitudal ensemble average is given as 
\begin{equation}
\label{eq:Magnitudal-Ensemble-Average}
\left<f{\left(\boldsymbol{\overline{N}}\right)}\right> = \frac{\sum\limits_{\boldsymbol{\overline{N}} \in \widetilde{C}{\left({M,N}\right)}}{f{\left(\boldsymbol{\overline{N}}\right)}\,\, g{\left(\boldsymbol{\overline{N}}\right)}}}{\sum\limits_{\boldsymbol{\overline{N}} \in \widetilde{C}{\left({M,N}\right)}}{g{\left(\boldsymbol{\overline{N}}\right)}}}\textrm{.}
\end{equation}
In the above equation, $\widetilde{C}{\left({M,N}\right)}$ is the set of each and every permutations of each and every partition of $N$ into $M$ parts (see Section~\ref{sec:multiplicity-function}). 
In column 4 of Table~\ref{tab:two-} the author provides the product of the multiplicity and the directional ensemble average of the dot product of the electoral-excess with itself. Thus, looking at the very last row of the table, the reader observes that its magnitudal ensemble average is $656.1/243 = 2.7$.

\newcommand{\vect}[1]{\boldsymbol{#1}}

\subsection{\label{sec:fractional-fluctuation}Fractional Fluctuation} 
The fractional fluctuation is a relative measurement of the expected deviation from equilibrium of a system. The present author gives the fractional fluctuation as the square root of the expected value of the inner product of the excess vector with itself, which is then normalized. It is mathematically defined in terms of the electoral vector, $ \boldsymbol{\overline{N}}$, the equilibrium-electoral vector, $ \boldsymbol{\overline{N}_o}$, and the one vector, $ \boldsymbol{\overline{1}}$, as
\begin{equation}
\label{eq:define-fractional-fluctuation}
\mathcal{F}{\left(N,M\right)} := \frac{\sqrt{\left<\left\{ { \left( \boldsymbol{\overline{N}} - \boldsymbol{\overline{N}_o}\right)} \boldsymbol{ \cdot} \left( \boldsymbol{\overline{N}} - \boldsymbol{\overline{N}_o}\right) \right\}\right>} }{\left< \left\{ {\boldsymbol{\overline{N}} \boldsymbol{ \cdot} \vect{\overline{1}}} \right\}\right>} \,\textrm{.} 
\end{equation}
This form above is suggestive of the coefficient of variation (i.e., the square root of the variance over the mean). Before continuing to explain how the fractional fluctuation is used to construct the four dynamic variables of interest, the author notes that the literature is without consensus on how to write the coefficient of variation for a multivariate distribution\cite{zhang2010multivariate, bennett1977multivariate}.

\subsection{\label{sec-four-dynamic-variables}Volatility, Flexibility, Stability, and Rigidity}
The technical approach is to write the fractional fluctuation in a separable form as
\begin{equation}
\label{eq:frac-fluc-separable}
\mathcal{F}{\left(N,M\right)} = {\mathcal{V}}{\left(N\right)}\,\mathcal{L}{\left(M\right)} \, ,
\end{equation}
where the volatility, $\mathcal{V}$, is solely a function of quantity of electors and the the flexibility, $\mathcal{L}$, is a function of the quantity of parties (and their probability of representation [see Section~\ref{sec:quantitative-results}]). The fractional fluctuation is proportional to both the volatility an the flexibility of the system. Meanwhile, the stability, $\mathcal{S}$, and rigidity, $\mathcal{R}$, of a system's properties are defined as the unitary complements of the volatility and fluidity, respectively, as
\begin{align}
\label{eq:stability-defined}
\mathcal{S}{\left(N\right)} & := 1 - \mathcal{V}{\left(N\right)} \textrm{, and}
\\
\label{eq:rigidity-defined}
\mathcal{R}{\left(M\right)} & := 1 -\mathcal{L}{\left(M\right)} \textrm{.}
\end{align}

Now that the technical approach is clarified, the author presents the results.

\section{\label{sec:results}Results}
In this section, the author presents two major results: the stability and the rigidity of the properties of electoral systems. The author presents a qualitative and quantitative explanation for both of these dynamic attributes.  The author begins with a qualitative description that may be gleaned through visual examination of the probability-density functions of the multinomially distributed random variables (Section~\ref{sec:qualitative-results}). The author ends with a quantitative description that is given by closed-form analytic expressions (Section~\ref{sec:quantitative-results}). 

\subsection{\label{sec:qualitative-results}Qualitative Results}
\begin{figure*}[hp!]
\begin{center}
\subfigure[Number electors equals 1, number parties equals 3.]{
\label{fig:pane-a}
\includegraphics[width=.47\textwidth]{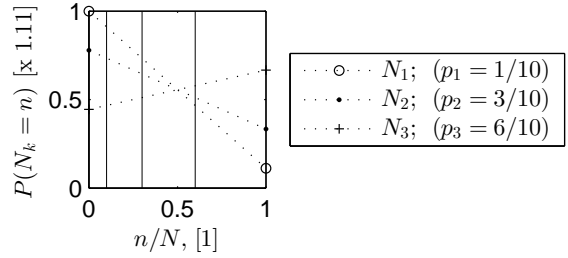}}\\
\vspace{-0.7em}
\subfigure[Number electors equals 6, number parties equals 3.]{
\label{fig:pane-c}
\includegraphics[width=.470\textwidth]{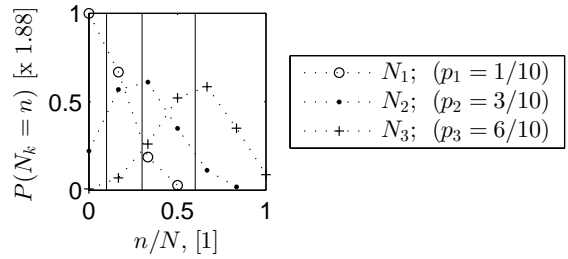}} \\
\vspace{-0.7em}
\subfigure[Number electors equals 15, number parties equals 3.]{
\label{fig:pane-e}
\includegraphics[width=.470\textwidth]{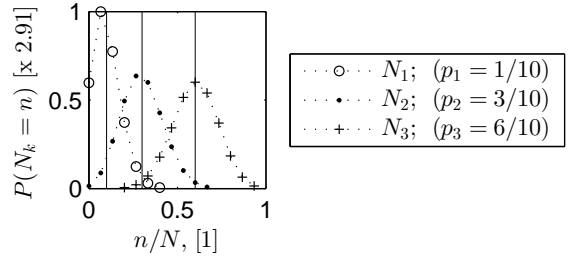}}\\
\vspace{-0.7em}
\subfigure[Number electors equals 200, number parties equals~3.]{
\label{fig:pane-i}
\includegraphics[width=.470\textwidth]{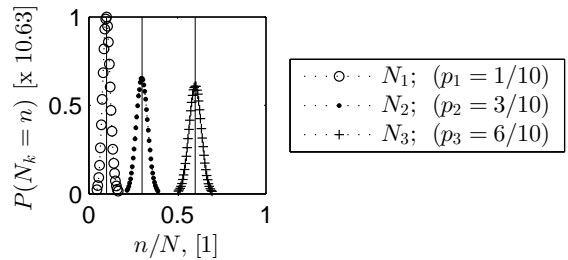}}\\
\caption{\label{fig:results-stability} Stability. The probability-density functions are plotted as a function of the normalized quantity of electors aligned with the $k^\textrm{th}$ outcome. Results in Panes (a), (b), (c), and (d) are parameterized with respect to the quantity of electors. Vertical lines are the quantity of electors at equilibrium.}
\end{center}%
\end{figure*}

\begin{figure*}[hp!]
\begin{center}
\subfigure[$\left(N,M\right) = \left(30,1\right)$]{
\label{fig:pane-b}
\includegraphics[width=.47\textwidth]{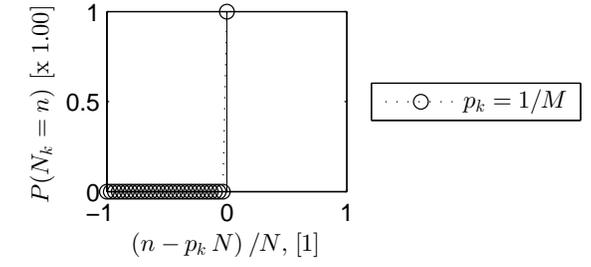}}\\
\vspace{-0.7em}
\subfigure[$\left(N,M\right) = \left(30,2\right)$]{
\label{fig:pane-d}
\includegraphics[width=.50\textwidth]{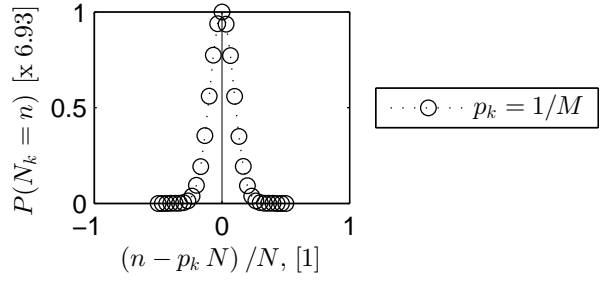}}\\
\vspace{-0.7em}
\subfigure[$\left(N,M\right) = \left(30,9\right)$]{
\label{fig:pane-f}
\includegraphics[width=.50\textwidth]{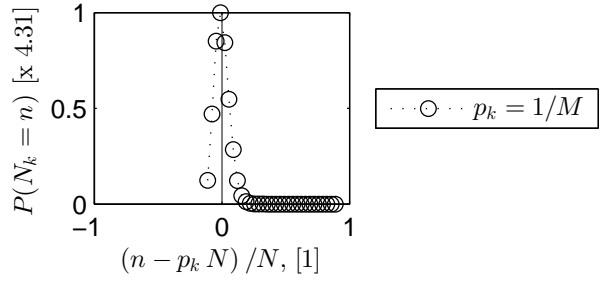}}\\
\vspace{-0.7em}
\subfigure[$\left(N,M\right) = \left(30,500\right)$]{
\label{fig:pane-j}
\includegraphics[width=.50\textwidth]{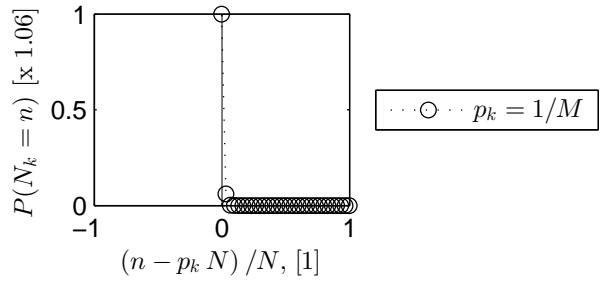}}\\
\caption{\label{fig:results-rigidity} Rigidity. The probability-density functions are plotted as a function of the normalized quantity of excess electors aligned with the $k^\textrm{th}$ outcome. Results in Panes (a), (b), (c), and (d) are parameterized with respect to the quantity of possible outcomes. Vertical lines are the electoral excess at equilibrium.}
\end{center}%
\end{figure*}

To illustrate the stability of electoral systems the author presents data for four distinct electoral systems in Figure~\ref{fig:results-stability}. In Panes (a), (b), (c), and (d), of Figure~\ref{fig:results-stability}, the quantity of parties is fixed at $M=3$ and the the probabilities for each outcome are fixed at  $p_1 =0.1$, $p_2 =0.3$, and $p_3 =0.6$. That said, the quantity of electors are $N=1$, $N=6$, $N=15$, and $N=200$, respectively. Similarly, to illustrate the rigidity of electoral systems the author presents data for four distinct electoral systems in Figure~\ref{fig:results-rigidity}. In Panes (a), (b), (c), and (d), of Figure~\ref{fig:results-rigidity} the quantity of electors is fixed at $N=30$ and the the probabilities for each and every outcome is uniform (i.e., $p_1 = p_2 = \cdots = p_M$). That said, the quantity of parties are $M=1$, $M=2$, $M=30$, and $M=500$, respectively. The $y$ axes of the eight plots are  the probability-density functions for each outcome.  For efficacy of viewing, in each pane the author scales the results. For example, in Pane~{(c)} of Figure~\ref{fig:results-stability}, each and every data point is multiplied by 2.91 before they are plotted.  The $x$ axes of the four plots in Figure~\ref{fig:results-stability} are identical.  On the $x$ axis the author present the quantity of electors aligned with each of the outcomes, $N_i$ (see~Equation~\ref{eq:electoral-vector}),which are all normalized by $N$.  The $x$ axes of the four plots in Figure~\ref{fig:results-rigidity} are identical.  On the $x$ axis the author present the quantity of excess electors aligned with each of the outcomes, $N_i$ (see~Equation~\ref{eq:electoral-vector}),which are all normalized by $N$. 


\subsubsection{Stability}
Just as in the binary system\cite{kittel-1980}, the stability of the system's properties increases as the multinomial distribution becomes more resolved and more steeply varying away from the peaks that are located at equilibrium. 

\paragraph{Resolution:}
In Figure~\ref{fig:results-stability}, the reader notes, that in all cases $N+1$ discrete points span the domain of $N_i/N$. Therefore, the resolution (in the optical sense) is coarser for smaller $N$ and finer for larger $N$. When the quantity of electors is few, such as in Pane (a), there are no trends or features to be readily observed. One may say that an electoral system with few electors are unpredictable to the observer. Conversely, as the quantity of electors increases (c.f., Panes (a) to (b) to (c) to (d)), the properties of the system are increasingly well resolved. One may say that as quantity of electors increases the properties of electoral system are increasingly predictable

\paragraph{Sharpness:}
Once the number of electors exceeds some minimum threshold, the reader observes that each outcome's distribution is uni-modal, with the mode of the $i^\textrm{th}$ outcome occurring at the abscissa value $N_i/N = p_i$ . Furthermore, as the reader successively views Panes (a), (b), (c), and (d),  the variance of each distribution gets smaller and the probability density sharply increases around the location of the equilibrium values (i.e., $N_i/N = N\,p_i/N = p_i$) and otherwise approaches null. Ultimately, in the limit as the quantity of parties approaches infinity the multiplicities approach  delta functions centered around their respective equilibria values. A larger variance indicates a system whose properties have  greater volatility. A smaller variance indicates a system whose properties have greater stability. Therefore, the properties of the electoral system become increasingly stable because the system is increasingly likely to be found closer and closer to equilibrium. 

\subsubsection{\label{sec:rigidity-explained}Rigidity}
The author's understanding of rigidity conforms to the normative usage\footnote{{\bf Rigid$\quad$1 a :} very firm rather than pliant in composition or structure {\bf:} lacking or devoid of flexibility $\ldots$ $\left<\textrm{a rigid totalitarian system -- Harrison Smith}\right>\dots\quad$ {\bf 6 : } of, relating to, or constituting a branch of dynamics in which the bodies whose motions are considered are treated as being absolutely invariable in shape and size under the application of force\cite{dic}.}. Before continuing, the reader is encouraged to bare in mind that the qualitative discussion herein applies specifically to the maximum-entropy, special case where $p_k = 1/M$ for each of the $k$ outcomes. A more general and thorough description of the rigidity of electoral systems is offered analytically in the quantitative results of Section~\ref{sec:quantitative-results}.

\paragraph{$\boldsymbol{M=1}$; Absolute Rigidity:}
In exactly the same way that Wolin writes that political rigidity is the absence of alternatives\cite{wolin}, the reader observes in Pane (a) of Figure~\ref{fig:results-rigidity} that there is one and only one value of $N_k$ with a non-zero probability density, namely the value at equilibrium (i.e., $N_k = \frac{1}{1}\,N$). Similarly, in exactly the same sense as the mechanics of a rigid body, a one-party system admits no fluctuations. Resulting from the fact that the electoral moments of each and every elector are the same, the system admits no variation in response to an electoral force -- the change in momentum of each and every elector is identical under the application of an electoral force.

\paragraph{$\boldsymbol{M=2}$; Two-Party System:}
The author plots the special case of an electoral system with two parties in Pane (b) of Figure~\ref{fig:results-rigidity}. For the case of $M=2$, the probability distribution is symmetric around the modal equilibrium value ($N_k = \frac{1}{2}\,N)$. As opposed to the results in Pane (a), there are now $N+1$ integers in the domain of Pane (b) that yield a non-zero probability density. In addition to the contribution towards the fractional fluctuation made by the value at equilibrium (i.e., $N_k = \frac{1}{2}\,N$), there are now an additional $N/2$ unique alternative values of the excess (i.e., $\pm 1, \pm 2, \ldots, \pm (N-2)/2, \pm N/2$) that make a contribution to the sum of squares (see Equation~\ref{eq:define-fractional-fluctuation}). 

\paragraph{$\boldsymbol{M}$ Finite; Multiparty System:}
For the sake of simplicity, in this paragraph the author considers $N_k$ to be a continuous real-valued random variable as opposed to a discrete random variable (i.e., $N_k \in \mathbb{R}$). The reader notes from the results given in Panes (b), (c), and (d) of Figure~\ref{fig:results-rigidity}, that as the quantity of parties increase from two towards infinity, that the probability distribution is increasingly less symmetric around the modal equilibrium value located ($N_k = \frac{1}{M}\,N)$. Thus, the domain of absolute values of the excess (i.e., $\left|n - p_k\,N \right| \in \left[0, \frac{M-1}{M} \,N \right]$) increases; and as it does there is a concomitant increase in the quantity of alternative contributions to the fractional fluctuation (i.e., $\left(n - p_k\,N \right)^2 \in \left[0, \left(\frac{M-1}{M}\right)^2 \,N^2 \right]$. Each of these additional alternatives presents additional modalities with which to respond to an electoral force.

\paragraph{Limit as $\boldsymbol{M}$ Approaches Infinity:}
In Pane (d) of Figure~\ref{fig:results-rigidity}, the author plots the probability-density function of the random variable $N_k$ for a very large quantity of outcomes $M= 500$. The reader notes that in the limit as the quantity of outcomes approaches infinity, the probability distribution increasingly appears as a reflected copy of the probability-density function for the one-party system in Pane (a) of Figure~\ref{fig:results-rigidity}. Indeed, as $M$ approaches infinity, $p_k$ goes towards $1/M$ and the variance of the binomially distributed variable goes as $\mathcal{O}{\left[\left<\left(N_k - p_k\,N\right)^2\right>\right]} = \mathcal{O}\left[N\,\frac{1}{M}\,\left(1 - \frac{1}{M}\right)\right] = \frac{N}{M} $. Thus just like in Pane (a),  as $M$ approaches infinity the variance of the random variable $N_k$ yields zero variability. However, as opposed to a one-party system there are now $M$ such binomially distributed random variables in the system. Consequently the total variance of the multinomially distributed electoral system asymptotically approaches $N$ (i.e., $M\, \frac{N}{M} = N$) as opposed to null.

\subsection{\label{sec:quantitative-results}Quantitative Results}
The author begins this section by presenting the closed-form expression for the fractional fluctuation. The interested reader may find a detailed proof of the expression in Appendix~\ref{sec:Proof-A}. It is then obvious to write analytic forms for the dynamic attributes of interest (Section~\ref{sec-four-dynamic-variables-reprieve}). The author ends this section by offering extremal values for the fractional fluctuation, flexibility and rigidity, all of which depend on the probability of representation of each and every possible outcome. This portion of the manuscript enables the reader to fully comprehend why the author chose to highlight the probability-density functions for the special case of $p_k = 1/M$ (see Figure~\ref{fig:results-rigidity} and its associated discussion in Section~\ref{sec:rigidity-explained}). The interested reader may find a detailed proof of the extremal values of these attributes in Appendix~\ref{sec:proof-B}.

\subsubsection{\label{sec-four-dynamic-variables-reprieve}Volatility, Flexibility, Stability, and Rigidity}
In Appendix~\ref{sec:Proof-A} the author derives the following expression for the fractional fluctuation. 
\begin{equation}
\label{eq:fractional-fluctuation-result}
\mathcal{F}{\left(N,M\right)} = \left[\frac{1}{\sqrt{N}}\,\right]\, \left[\sqrt{1 - \sum_{i=1}^M{p_i^2} }\,\,\,\right] \textrm{.}
\end{equation}
Comparing the result above with the separable form of the fractional fluctuation that is given in Equation~\ref{eq:frac-fluc-separable}, the volatility, $\mathcal{V}{\left(N\right)}$, and flexibility, $\mathcal{L}{\left(M\right)}$, of the electoral system are respectively revealed by the following closed-form analytic expressions:
\begin{align}
\mathcal{V}{\left(N\right)} & = \frac{1}{\sqrt{N} } \, \textrm{, and}
\\
\mathcal{L}{\left(M\right)} & = \sqrt{ 1 - \sum_{i=1}^M{p_i^2} } \, \textrm{.}
\end{align}
Next, comparing the results above with the definitions in Equations~\ref{eq:stability-defined} and \ref{eq:rigidity-defined}, the stability, $\mathcal{S}{\left(N\right)}$, and rigidity, $\mathcal{R}{\left(M\right)}$, of the electoral system are respectively revealed by the following closed-form analytic expressions:
\begin{align}
\label{eq:stability-closed-form}
\mathcal{S}{\left(N\right)} & = 1 - \frac{1}{\sqrt{N} } \, 
\textrm{, and}
\\
\label{eq:rigidity-closed-form}
\mathcal{R}{\left(M\right)} & = 1- \sqrt{ 1 - \sum_{i=1}^M{p_i^2} } \, \textrm{.}
\end{align}
As explained in the technical approach (see Section\ref{sec-four-dynamic-variables}), the volatility and stability of the system's properties are solely dependent on the quantity of electors while flexibility and rigidity of the system's properties are dependent on the quantity of parties (and each party's probability of representation).

\subsubsection{\label{sec:extremal-values}Extremal Values}
Beginning with absolute instability (i.e., $\mathcal{S}{\left(1\right)} = 0$) for a single-elector system, the stability increases monotonically to unity in the limit as the quantity of electors approaches infinity (i.e., $\lim_{M \rightarrow \infty}\mathcal{S}{\left(M\right)} = 1$).

In Appendix~\ref{sec:proof-B}, the author shows that for a fixed integer $M$ of possible outcomes the flexibility is bounded as
\begin{equation}
\label{eq:bounds-flexibility}
0 \, \leq \mathcal{L}{\left(M\right)} \leq  \sqrt{\frac{M - 1 }{M} } \textrm{.}
\end{equation}
The flexibility has a maximum when each of the   $M$ of possible outcomes is equally probable, and has a minimum when any one of the possible outcomes is one-hundred percent probable.  The rigidity is thus bounded as 
\begin{equation}
1 - \sqrt{\frac{M-1}{M} } \leq \, \mathcal{R}{\left(M\right)} \leq 1 \textrm{.} 
\end{equation}
The rigidity has a maximum when any one of the possible outcomes is one-hundred percent probable, and has a minimum when each of the $M$ possible outcomes is equally probable. 
The properties of an electoral system with one possible outcome are absolutely rigid.  As the quantity of possible outcomes increases and the respective probabilities of each outcome are increasingly proporional, then the properties of the electoral system become less rigid and more flexible. 

\section{\label{sec:discussion}Discussion}
Until now the author offers mathematical notions of electoral dynamics.  As a counterpoint, the author now discusses how this theory is relevant to realpolitik.  The author asks and answers the following question: how does the theory of electoral stability and rigidity enhance our understanding of the constituional order of the United States of America, which has multiple branches of government?  An individual analysis of each of three branches of government is presented in Section~\ref{sec:individual-analysis}.  A comparitive analysis of three branches of government is presented in Section~\ref{sec:comparative-analysis}.  A tabular  summary  is given in Table \ref{tab:multiply-branched-systems}.
\begin{table}[hbt!]
\begin{center}
\renewcommand*{\arraystretch}{1.4}
\caption{\label{tab:multiply-branched-systems}Realpolitik. Summary of the individual analyses of the governmental branches of the  United States of America. The brackets indicate a range of values.}
\vspace{0.3em}
\begin{tabular}{| c| c| c|}
\hline
Governmental Branch & Stability & Rigidity \\ 
\hline
\hline
President / Vice-President & $\left[0.00, 0.29\right]$ & $\left[0.00, 1.00\right]$ \\
\hline
Supreme Court & $\left[0.59, 0.67\right]$ & $\left[0.00, 1.00\right]$ \\
\hline
Senate &  $\left[0.86, 0.90\right]$ & $\left[0.01, 1.00\right]$ \\
\hline
House of Representatives & $\left[0.93, 0.95\right]$ & $\left[0.00, 1.00\right]$ \\
\hline
\end{tabular}
\end{center}
\end{table}

\subsection{\label{sec:individual-analysis}Individual Analysis}
An individual analysis of the executive branch, the judicial branch, and the legislative branch is given in Sections~\ref{sec:executive-ranch}, \ref{sec:judicial-ranch}, and 
\ref{sec:legislative-branch}, respectively.  The stability and rigidity of each branch is sequentially  described.  Each attribute is bounded from below and bounded from above. 

\subsubsection{\label{sec:executive-ranch}Executive Branch}
The executive branch of government is  comprised of the federal offices of the President and the Vice President.  If the executive branch only included the president, then the executive branch would be a single-elector system and thus absolutely unstable (i.e., $\mathcal{S}{\left(1\right)} = 0$). Because the founding fathers incorporated a line of succession\footnote{U.S. Constitution, Article II, Section 1, Clause 5.} that includes the federal office of the Vice President, the executive branch is stabilized -- at least to some extent.  In consideration of these facts, the author states that the stability of the executive branch, $\mathcal{S}_E$, is bounded as $\mathcal{S}{\left(1\right)}  \leq \mathcal{S}_E  \leq \mathcal{S}{\left(2\right)}$.

When deliberating any choice,  the President may seek extensive council and demand multiple options from advisors and  administrators.  Ultimately the buck passing stops with the President who is solely responsible  to reduce the quantity of choices  and select an outcome.  Each choice made by the President is circumstantial.   At any time, there may be  a myriad of choices under deliberation.  In consideration of these facts, the author states that the rigidity of the executive branch, $\mathcal{R}_E$, is bounded as $\mathcal{R}{\left(\infty\right)}  \leq \mathcal{R}_E  \leq \mathcal{R}{\left(1\right)}$.

\subsubsection{\label{sec:judicial-ranch}Judicial Branch}The judicial branch of government is comprised of the federal offices of the Justices of the Supreme Court. Every decision is made by a majority vote; however, justices may recuse themselves from participation in any official action. That said, six members of the court constitute a quorum. In consideration of these facts, the author states that the stability of the judicial branch, $S_J$, is bounded as $\mathcal{S}{\left(6\right)} \leq \mathcal{S}_J \leq \mathcal{S}{\left(9\right)}$. In theory, lady justice is blind and so non-partisan.  The justices, in their robes may be considered either indistinguishable from each other or disitinguishable (i.e., independent judicial review). Subsequently, there are often unanimous decision and there are also often more than one dissenting opinion.  In consideration of these facts, the author states that the rigidity of the judical branch, $R_J$, is bounded as $\mathcal{R}{\left(9\right)} \leq \mathcal{R}_J \leq \mathcal{R}{\left(1\right)}$. 

\subsubsection{\label{sec:legislative-branch}Legislative Branch}
The legislative branch of government is bicameral; it is composed  of the Senate and the House of Representatives. The quantity of electors in these branches are 101 and 430, respectively.  Article~I, Section~5 of the Constitution requires that a quorum to conduct official Senate business is the smallest integer larger than half the quantity of electors (i.e. 51). Whereas given by the clerk of the House of Representatives, ``when there are no vacancies in the membership, a quorum is 218.\footnote{http://clerk.house.gov/legislative/legfaq.aspx}.'' In consideration of these facts, the author states that the stability of the Senate, $S_S$, and the House of Representatives, $S_R$, are bounded as $\mathcal{S}{\left(51\right)} \leq \mathcal{S}_S \leq \mathcal{S}{\left(101\right)}$ and $\mathcal{S}{\left(218\right)} \leq \mathcal{S}_R \leq \mathcal{S}{\left(430\right)}$, respectively. 

With respect to rigidity, while party cohesion is certainly factored into their choices, each member of congress is  also tied to  the parochial interests of their local constituency, as well as the political patronage from within and without their respective constituency. Therefore, as in the judical branch, the author states that the rigidity of each camera is bounded from below by their respecive quantity of members (i.e., $\mathcal{R}{\left(M=N\right)}$); and is bounded from above by the possibility of the electors acting in unison with consensus (i.e., $\mathcal{R}{\left(M=1\right)}$). In consideration of the above, author states that the rigidity of the Senate, $R_S$, and the House of Representatives, $R_R$, are bounded as $\mathcal{R}{\left(101\right)} \leq \mathcal{R}_S \leq \mathcal{R}{\left(1\right)}$ and $\mathcal{R}{\left(430\right)} \leq \mathcal{R}_R \leq \mathcal{R}{\left(1\right)}$, respectively.

\subsection{\label{sec:comparative-analysis}Comparative Analysis}
The summary of individual analyses of each branch of government is given in Table~\ref{tab:multiply-branched-systems}. The  data is consistent with the experience of the everyday lay political observer and the astute specialist. The author finds that resulting from the inherent instability of the executive branch  it is no wonder that the election of the President/Vice President is the most watched and the most contested of all. Further, one recognizes what the founders tacitly understood; specifically that a succession plan acts to stabilize the executive branch in case of a national tragedy or an  untimely death. The second most fevered pitch in the political calendar of the United States occurs when the nation undergoes a Supreme-Court confirmation. It is precisely because of the relatively low stability of a nine-seat jury opining on  issues fundamently relevent to 300 million  people that the justices be thoroughly vetted  and even subject to a super-majority vote. To first-order approximation, among the bicameral legislative branches, the  observer notes that the Senate, with its lesser quantity of electors and concomitant greater instability, is the more valued legislative prize for the parties that comprise the political order; and that  the House of Representative, with its greater quantity of electors and greater stability, provides the  highest fidelity representation of the people. 

\section{\label{sec:conclusions}Conclusions}

Given that the number of electors is at most 8 billion () we may not safely presume that the most probably configuration is the only configuration is the only significant configuration for electoral systems.

This manuscript is inspired by the Supreme Court's majority opinion on Timmons versus the Twin-Cities Area New Party. The author draws on techniques well known in statistical mechanics to refute the majority opinion of the court. In conclusion, (\emph{i}) electoral stability and electoral rigidity are best served by increasing the quantity of electors, (\emph{ii}) electoral rigidity is best served by  decreasing the quantity of parties and reducing their proportional representation, and  (\emph{iii}) the less stable a branch of government, the more concern is placed on those who would hold those offices for the people.


\begin{thebibliography}{10}

\bibitem{renquist-1997}
{Timmons v. Twin Cities Area New Party}, 520 U.S. (1997). 351.

\bibitem{black-1958}
D.~Black, {\it The Theory of Committees and Elections\/}, I.~McLeon,
  A.~McMillan, B.~L. Monroe, eds., Revised Second Editions (Kluwer Academic
  Publishers, 1998), pp. 55--102, 189--270.

\bibitem{borda-1781}
M.~de~Borda, M\'{e}moire sur les \'{E}lections au {S}crutin (1781).

\bibitem{condorcet-1785}
M.~de~Condorcet, {\it Essai sur l'Appliation de l'Analyse \'{a} la
  Probilit\'{e} dex D\'{e}citions Rendues a la Plualit\'{e} de Voix\/}
  (Imprimerie Royale, 1785).

\bibitem{laplace-1812}
P.-S.~M. de~LaPlace, {\it Journal de l'Ecole Polytechnique\/} {\bf tome II}
  (1812).

\bibitem{dodgson-1873}
C.~L. Dodgson, A discussion of the various methods of procedure in conducting
  elections, Pamphlet (1873).

\bibitem{dodgson-1874}
C.~L. Dodgson, Suggestions as to the best method of taking votes, where more
  than two issues are to be voted on, Pamphlet (1874).

\bibitem{dodgson-1876}
C.~L. Dodgson, A method of taking votes on more than two issues, Pamphlet
  (1876).

\bibitem{nanson-1882}
E.~Nanson, {\it Proceedings of the Royal Society of Victoria\/} (1882),
  vol.~19, pp. 197--240.

\bibitem{dodgson-1884}
C.~L. Dodgson, {\it The Principles of Parliamentary Representation\/} (Harrison
  \& Sons, 1884).

\bibitem{hoag-1926}
C.~G. Hoag, G.~H. Hallet, {\it Proportional Representation\/} (Macmillan,
  1926).

\bibitem{arrow-1951}
K.~Arrow, {\it Social Choice and Individual Values\/} (John Wiley \& Sons,
  1963), second edn.

\bibitem{key-1955}
{V. O. Key, Jr.}, {\it The Journal of Politics\/} {\bf 17}, 3 (1955).

\bibitem{campbell-1966}
A.~Campbell, {\it Elections and the Political Order\/} (John Wiley and Sons,
  Inc., 1966).

\bibitem{ockham-1300}
W.~Ockham, {\it Philosophical Writings: A Selection\/} (Bobbs-Merrill Company,
  Inc., 1964), pp. xx--xxi.

\bibitem{warner-1880}
C.~D. Warner, {\it Philosophical Writings: A Selection\/} (Sampson Low,
  Marston, Searle, \& Rivington, 1880), pp. xx--xxi.

\bibitem{farrier2010congressional}
J.~Farrier, {\it Congressional Ambivalence: The Political Burdens of
  Constitutional Authority\/} (University Press of Kentucky, 2010).

\bibitem{craig2005ambivalence}
S.~Craig, M.~Martinez, {\it Ambivalence, Politics and Public Policy\/}
  (Palgrave Macmillan US, 2005).

\bibitem{kittel-1980}
C.~Kittel, H.~Kroemer, {\it Thermal Physics\/} (W. H. Freeman and Company,
  1980), pp. 10--24, second edn.

\bibitem{leibniz}
G.~W. Leibniz. Math. Schriften, Vol. IV 2, Specimen de divulsionibus
  aequationum . . . Letter 3 dated Sept. 2, 1674.

\bibitem{euler}
L.~Euler, {\it Introduction to Analysis of the Infinite\/} (Springer, 1988).
  Translated by J. Blanton.

\bibitem{zhang2010multivariate}
L.~Zhang, {\it et~al.\/}, {\it Accreditation and Quality Assurance\/} {\bf 15},
  351 (2010).

\bibitem{bennett1977multivariate}
B.~Bennett, {\it Statistische Hefte\/} {\bf 18}, 123 (1977).

\bibitem{dic}
P.~Gove, ed., {\it Wester's Third New International Dictionary of the English
  Language Unabridged\/} (Merriam-Webster Inc., 1993).

\bibitem{wolin}
S.~Wolin, {\it Democracy Incorporated: Managed Democracy and the Specter of
  Inverted Totalitariansim\/} (Princeton University Press, 2010), p.~xv, fifth
  printing and first paperback printing edn.

\bibitem{crc}
D.~Zwillinger, ed., {\it Standard Mathematical Tables and Formulae\/} (CRC
  Press, 1996), pp. 170, 441, 582--3, 30th edn.

\end{thebibliography}


\appendix

\section{\label{sec:Proof-A}Proof of Equation~\ref{eq:fractional-fluctuation-result}: The Fractional Fluctuation}
The averaged expected value of the inner product of the electoral-excess vector with itself for an isotropic moment is expanded as given as

\subsection{Dual Ensemble Average}
The fractional fluctuation is given as the dual-ensmeble average of the dot product of the electoral-excess vector with itself (compare Equation~\ref{eq:electoral-excess-vector} and Equation~\ref{eq:define-fractional-fluctuation}). This dot product may be expansed as
\begin{align}
\boldsymbol{\overline{x}}\cdot \boldsymbol{\overline{x}} 
&
=
\left[ \sum_{k=1}^{M} {\left( x_k \, \boldsymbol{\hat{k}} \right)} \right] \boldsymbol{\cdot} \left[ \sum_{l=1}^{M} \left(x_l\,\boldsymbol{\hat{l}} \right)\right]
\\
&
=
\sum_{k=1}^{M} \sum_{l=1}^{M} {\left( x_k \, \boldsymbol{\hat{k}} \right) \boldsymbol{\cdot} \left(x_l\,\boldsymbol{\hat{l}} \right)} 
\\
&
= \sum_{k=1}^{M} \sum_{l=1}^{M} {\left(x_k\,x_l\right)\, \left(\boldsymbol{\hat{k}} \boldsymbol{\cdot} \boldsymbol{\hat{l}}\right)} 
\\
&
= 
\sum_{k=1}^{M} {x_k^2} + \sum_{k=1}^{M} \sum_{l=1}^{M} {\left(x_k\,x_l\right)\, \left(\boldsymbol{\hat{k}} \boldsymbol{\cdot} \boldsymbol{\hat{l}}\right)} \,\left(1 - \delta_{k,l}\right) \\
\textrm{.}
\end{align}
The Kronecker delta function, $\delta_{k,l}$, is equal to zero when $k \neq l $ and is one otherwise; $x_k = N_k - p_k\,N$.

Next we consider the two distinct averages (see Equation~\ref{eq:Directional-Ensemble-Average} and Equation~\ref{eq:Magnitudal-Ensemble-Average}). A magnitudal average, $\left< \boldsymbol{\overline{x}}\cdot \boldsymbol{\overline{x}} \right>$, which averages $x_k\,x_l$ over each and every of the $M^N$ allowed states of the system; and a directional average, $\left\{ \boldsymbol{\overline{x}}\cdot \boldsymbol{\overline{x}} \right\}$, which averages $\boldsymbol{\hat{k}} \boldsymbol{\cdot} \boldsymbol{\hat{l}}$ over each and every allowable angles subtended between them. The derivation of the former is rather straitforward and depends on known properties of the multinomial distribution. The derivation of the latter is rather more involved; and so while the result is given below, the interested reader may view the entire derivation in Section~\ref{sec:directional-ensemble-derivation}. The author finds that 
\begin{align}
\left< \boldsymbol{\overline{x}}\cdot \boldsymbol{\overline{x}} \right>
=
&
\sum_{k=1}^{M} {\left< x_k^2 \right> } + \sum_{k=1}^{M} \sum_{l=1}^{M} {\left<x_k\,x_l\right>\, \left( \boldsymbol{\hat{k}} \boldsymbol{\cdot} \boldsymbol{\hat{l}} \right)} 
\\
=
&
\sum_{k=1}^{M} { N \, p_k \,\left(1- p_k\right) } - \sum_{k=1}^{M} \sum_{l=1}^{M} {N\,p_k\,p_l\, \left( \boldsymbol{\hat{k}} \boldsymbol{\cdot} \boldsymbol{\hat{l}} \right)} 
\\
\label{eq:average-1}
=
&
N \left[ 1 - \sum_{k=1}^{M} { p_k^2 } - \sum_{k=1}^{M} \sum_{l=1}^{M} {p_k\,p_l\, \left( \boldsymbol{\hat{k}} \boldsymbol{\cdot} \boldsymbol{\hat{l}} \right)} \right] 
\textrm{.}
\end{align}
\begin{align}
\left\{ \boldsymbol{\overline{x}}\cdot \boldsymbol{\overline{x}} \right\}
=
&
\sum_{k=1}^{M} {x_k^2} + \sum_{k=1}^{M} \sum_{l=1}^{M} {\left(x_k\,x_l\right)\, \left\{ \boldsymbol{\hat{k}} \boldsymbol{\cdot} \boldsymbol{\hat{l}} \right\}\,\left(1 - \delta_{k,l}\right)} 
\\ 
\label{eq:average-2}
= & \sum_{k=1}^{M} {x_k^2} 
\, \textrm{.}
\end{align}
With the use of Equation~\ref{eq:result-222}, the author writes the dual ensemble average as
\begin{align}
\label{eq:expected}
\left\{\left< \boldsymbol{\overline{x}}\cdot \boldsymbol{\overline{x}} \right> \right\} = \left<\left\{ \boldsymbol{\overline{x}}\cdot \boldsymbol{\overline{x}} \right\} \right> 
=
N \left[ 1 - \sum_{k=1}^{M} { p_k^2 } \right] 
\textrm{.}
\end{align}
A similar and straitforward derivation may be used to write the dot product of the excess vector with the one vector. The result being that 
\begin{align}
\label{eq:expected-first-order}
\left\{\left< \boldsymbol{\overline{x}}\cdot \boldsymbol{\overline{1}} \right> \right\} = \left<\left\{ \boldsymbol{\overline{x}}\cdot \boldsymbol{\overline{1}} \right\} \right> 
=
N\, \textrm{.}
\end{align}
From the above two equations, one obtain that which was to be demonstrated, namely the analytic form of the fractional fluctuation that is written in Equation~\ref{eq:fractional-fluctuation-result}.

\subsection{\label{sec:directional-ensemble-derivation}Directional Ensemble Average of Dot Product of Two Vectors}
In this section, the author allows that vectors $\boldsymbol{\hat{u}}_1, \boldsymbol{\hat{u}}_2, \ldots, \boldsymbol{\hat{u}}_M$ form the basis of a $M$-dimensional electoral vector space. Let $\boldsymbol{\hat{k}}$ be a radial unit vector. Allow that $\boldsymbol{\hat{k}}$ is isotropic and point in each and any direction on an $M$-sphere\footnote{Typically, this is called an $N$-sphere. To conform with the present nomenclature, here it is called an $M$-sphere.} with equal probability. The probability-density function for picking a vector $\boldsymbol{\hat{k}}$ with coordinates ($\phi_1, \ldots, \phi_{M-1}$) is 
\begin{equation}
\label{eq:Z}
p\left(\phi_1, \ldots, \phi_{M-1}\right) = \int_{\phi_{M-1}}^{2\,\pi}\,\int_{\phi_{M-1}}^{\pi}\, \cdots \, \int_{\phi_1}^{\pi} d^{M-1}S \, \textrm{,}
\end{equation}
where $d^{M-1}S$ is the volume element of an ($M-1$)-sphere or alternatively the surface element of an $M$-sphere. The surface element is written as
\begin{equation}
\label{eq:differential-area}
d^{M-1}S = \sin^{M-2}{\left(\phi_1\right)} \, \sin^{M-3}{\left(\phi_2\right)} \cdots \sin{\left(\phi_{M-2}\right)} \, d\phi_1\, d\phi_2 \cdots \, d\phi_{M-1} \textrm{.}
\end{equation}
The solution to the integral in Equation~\ref{eq:Z}
is well known as is given in terms of the gamma function, $\Gamma$, as
\begin{equation}
p\left(\phi_1, \ldots, \phi_{M-1}\right) = \frac{2\,\pi^{\left(M+1\right)/2}}{\Gamma{\left(\frac{M}{2}+1\right)}}
\end{equation}

Next, let $\boldsymbol{\hat{k}}$ and $\boldsymbol{\hat{l}}$ be two radial unit vectors. Let $\boldsymbol{\hat{k}}$ be fixed. Without loss of generality let 
\begin{equation}
\label{eq:X}
\boldsymbol{\hat{k}} = \boldsymbol{\hat{u}}_1 \textrm{.}
\end{equation}
In spherical coordinates, $\boldsymbol{\hat{k}}$ corresponds to the unit vector with $r=1$, $\phi_1 = 0$. Note that $\phi_2, \ldots, \phi_{M-1}$ are arbitrary. For the special case where $\boldsymbol{\hat{l}} = \boldsymbol{\hat{k}}$, the expected value $\left\{ \boldsymbol{\hat{k}} \boldsymbol{\cdot} \boldsymbol{\hat{l}} \right\}$ is given trivially as 
\begin{equation}
\label{eq:result-1}
\left.\left\{ \boldsymbol{\hat{k}} \boldsymbol{\cdot} \boldsymbol{\hat{l}} \right\}\right|_{ \boldsymbol{\hat{k}} = \boldsymbol{\hat{l}} } = \frac{ \int_{\phi_{M-1}}^{2\,\pi}\,\int_{\phi_{M-1}}^{\pi}\, \cdots \, \int_{\phi_1}^{\pi} 1\, d^{M-1}S}{\frac{2\,\pi^{\left(M+1\right)/2}}{\Gamma{\left(\frac{M}{2}+1\right)}}} = 1
\end{equation}
Next, let $\boldsymbol{\hat{l}}$ be isotropic and point in each and any direction with equal probability. Generally, in Cartesian coordinates 
\begin{equation}
\label{eq:Y}
\boldsymbol{\hat{l}} = \cos{\left(\phi_1\right)}\,\boldsymbol{\hat{u}}_1 + \sin{\left(\phi_1\right)}\,\cos{\left(\phi_2\right)}\,\boldsymbol{\hat{u}}_2 + \ldots +\sin{\left(\phi_1\right)}\,\sin{\left(\phi_2\right)}\,\ldots\sin{\left(\phi_{M-2}\right)}\, \sin{\left(\phi_{M-1}\right)} \, \boldsymbol{\hat{u}}_M\textrm{.}
\end{equation}
In spherical coordinates, $\boldsymbol{\hat{l}}$ corresponds to the unit vector with $r=1$, and $\phi_1, \ldots \phi_{M-1}$ are all arbitrary. So, in light of Equations~\ref{eq:X} and \ref{eq:Y}, 
the dot product $\boldsymbol{\hat{k}} \boldsymbol{\cdot} \boldsymbol{\hat{l}} = \cos{\left(\phi_1\right)}$. The expected value $\left\{ \boldsymbol{\hat{k}} \boldsymbol{\cdot} \boldsymbol{\hat{l}} \right\}$ is given as
\begin{equation}
\label{eq:eval-iso}
\left.\left\{ \boldsymbol{\hat{k}} \boldsymbol{\cdot} \boldsymbol{\hat{l}} \right\} \right|_{\boldsymbol{\hat{l}} \textrm{ isotropic}} = \frac{ \int_{\phi_{M-1}}^{2\,\pi}\,\int_{\phi_{M-1}}^{\pi}\, \cdots \, \int_{\phi_1}^{\pi} \cos{\left(\phi_1\right)} \, d^{M-1}S}{\frac{2\,\pi^{\left(M+1\right)/2}}{\Gamma{\left(\frac{M}{2}+1\right)}}} \,\textrm{.}
\end{equation}
Next, comparing Equations~\ref{eq:differential-area} and \ref{eq:eval-iso}; and bearing in mind that 
\begin{equation}
\int_{\phi_1 = 0}^{\pi}{ \cos{\left(\phi_1\right)} \, \sin^{M-2}{\left(\phi_1\right)} \, d\phi_1} = 0
\end{equation}
one obtains
\begin{equation}
\label{eq:result-222}
\left.\left\{ \boldsymbol{\hat{k}} \boldsymbol{\cdot} \boldsymbol{\hat{l}} \right\} \right|_{\boldsymbol{\hat{l}} \textrm{ isotropic}} = 0
\end{equation}

\section{\label{sec:proof-B}Proof of Equation~\ref{eq:bounds-flexibility}: The Bounds of Flexibility}
From Equation\ref{eq:rigidity-closed-form}, the flexibility  is. 
\begin{equation}
\label{eq:fluidity-fluctuation-result-rewrite}
\mathcal{L}{\left(p_1,\ldots, p_M\right)} = \sqrt{1 - \sum_{i=1}^M{p_i^2} }  \textrm{.}
\end{equation}
The extremal values of $\mathcal{L}$ occurs at the endpoint(s) and/or when the gradient of $\mathcal{L}$, $ \nabla{\mathcal{L}}$, is equal to zero.  Here, the author compares the values of the flexibility at the end point and the zero-gradient point and determines the minimum and maximum values of the flexibility.

The end point(s) occurs when $p_j = 1$ and $p_k = 0$ for all $k \in 1, \ldots, M$ and $k \neq j$ .  The value of the flexibility at the endpoint(s), $\left.\mathcal{L}\right|_{e.p.}$, is
\begin{equation}
\left.\mathcal{L}\right|_{e.p.} =   \sqrt{1 -1}  = 0 \textrm{.}
 \end{equation}
Meanwhile, the gradient is generally given\cite{crc} by 
\begin{align}
\nabla{\mathcal{L}} 
&
= \sum_{k=1}^M{ \left(\frac{\partial \mathcal{L}}{\partial p_k }\,\mathbf{\hat{k}}\right) }\,.
\end{align}
Plugging in from Equation~\ref{eq:partial}, the gradient is
\begin{align}
\nabla{\mathcal{L}} 
&
= \sum_{k=1}^{M-1}{\frac{\left(p_M  - p_k\right) \, \mathbf{\hat{k}} }{ \left[1 - \sum_{i=1}^M{p_i^2} \right]^{1/2}}}\,.
\end{align}
Thus for an arbitrary set of unit vectors, the gradient has a value of zero when $p_M = p_k$ for all $k \in 1, \ldots, M$.   Further, considering the constraint given in Equation~\ref{eq:contraint-equation}, the gradient has a value of zero when $p_k = 1/M$ for all $k \in 1, \ldots, M$. So the value of the flexibility when the gradient is zero, $\left.\mathcal{L}\right|_{z.g.p.}$, is
\begin{equation}
\left.\mathcal{L}\right|_{z.g.p.} = \sqrt{1 - \sum_{i=1}^M{\left(\frac{1}{M}\right)^2} } = \sqrt{\frac{M-1}{M}}  \textrm{.}
\end{equation}
The author finds  that which was to be demonstrated.  Namely, the flexibility has a maximum when each of the fixed integer $M$ of possible outcomes is equally probable and the flexibility has a minimum when any one of the possible outcomes is one-hundred percent probable.

\subsection{Constrained Rates of Change}
In the present of the constrain given in Equation~\ref{eq:contraint-equation}, we are allowed to talk about the partial derivatives of $\mathcal{L}$ only if $\mathcal{L}$ is expressed as a function of independent variables.  Without loss of generality, the author chooses the variables $p_1, \ldots, p_{M-1}$ to be the independent varbiables, and view $\mathcal{L}$ and $p_M$ as functions of the independent variables before talking about the partial derivatives.  The notation 
\begin{equation}
\left(\frac{\partial{\mathcal{L}}}{\partial{p_k}}\right)_{p_1, \ldots, p_{M-1}}
\end{equation}
means that we are viewing $\mathcal{L}$ as a function of the independent variables $p_1, \ldots, p_{M-1}$, and measuring the rate of change of $\mathcal{L}$ as $p_k$ ($k \in 1, \ldots, M-1$) varies while holding $p_1, \ldots, p_{k-1},  p_{k+1}, \ldots, p_{M-1}$ constant.

To compute  $\left(\frac{\partial{\mathcal{L}}}{\partial{p_1}}\right)_{p_1, \ldots, p_{M-1}}$, we need to express the differential flexibility, $d\mathcal{L}$, in the form
\begin{equation}
d\mathcal{L} = \sum_{k=1}^{M-1}{\left[dp_k\, \left(\frac{\partial{\mathcal{L}}}{\partial{p_k}}\right)_{p_1, \ldots, p_{M-1}}\right]} \,.
\end{equation}

Yet $\mathcal{L}$ is intially a function of the variables $p_1,\ldots, p_{M}$.  If $\mathcal{L} = \sqrt{1 - \sum_{p=1}^{M}{p_i^2}}$ is viewed as a function of $\mathbb{R}^M$, the definition of $d\mathcal{L}$ gives
\begin{align}
d\mathcal{L} &
 = 
\label{eq:q23432} 
\sum_{k=1}^{M}{\left[ \frac{-{p_k  }}{ \left[1 - \sum_{i=1}^M{p_i^2} \right]^{1/2}} \, dp_k \right]}.
\end{align}
If $\mathcal{L}$ is restricted to a function on the domain defined by the constraint  given in Equation~\ref{eq:contraint-equation}, then Equation~\ref{eq:q23432}  still holds.  Taking the differential of the constraint eqauation $1 = \sum_{k=1}^M{p_k}$ gives a relation between $p_1, \ldots, p_{M}$, which is
\begin{equation}
\label{eq:1123423432}
0 = \sum_{k=1}^{M}{ dp_k}.
\end{equation}
Because we want $d\mathcal{L}$ in terms of the $dp_1$, $\ldots$, $dp_{M-1}$ only, we solve Equation~\ref{eq:1123423432} for $dp_M$,
\begin{equation}
\label{eq:23423432}
 dp_{M} = - \sum_{j=1}^{M-1}{ dp_j }
\end{equation}
and substitute into Eqation~\ref{eq:q23432}, which yields that 
\begin{align}
d\mathcal{L} &
 = \sum_{k=1}^{M}{\left[dp_k\,\, \left(\frac{\partial{\mathcal{L}}}{\partial{p_k}}\right)_{p_1, \ldots, p_{M-1}}\right]}
\\
&
= 
\sum_{k=1}^{M}{ dp_k \, \frac{-{p_k  }}{ \left[1 - \sum_{i=1}^M{p_i^2} \right]^{1/2}}}
\\
&
= 
 { dp_M \, \frac{-{p_M  }}{ \left[1 - \sum_{i=1}^M{p_i^2} \right]^{1/2}}} + \sum_{k=1}^{M-1}{ dp_k \, \frac{-{p_k  }}{ \left[1 - \sum_{i=1}^M{p_i^2} \right]^{1/2}}} 
\\
&
= 
 {- \sum_{k=1}^{M-1}{ dp_k }\, \frac{-{p_M  }}{ \left[1 - \sum_{i=1}^M{p_i^2} \right]^{1/2}}} + \sum_{k=1}^{M-1}{ dp_k \, \frac{-{p_k  }}{ \left[1 - \sum_{i=1}^M{p_i^2} \right]^{1/2}}} 
\\
&
\label{eq:q23432121212}
= 
 \sum_{k=1}^{M-1}{ dp_k \, \frac{p_M  - p_k }{ \left[1 - \sum_{i=1}^M{p_i^2} \right]^{1/2}}} .
\end{align}

This means that 
\begin{equation}
\label{eq:partial}
\left(\frac{\partial{\mathcal{L}}}{\partial{p_k}}\right)_{p_1, \ldots, p_{M-1}} =  \frac{p_M  - p_k }{ \left[1 - \sum_{i=1}^M{p_i^2} \right]^{1/2}} .
\end{equation}
So the value of $\left(\frac{\partial{\mathcal{L}}}{\partial{p_k}}\right)_{p_1, \ldots, p_{M-1}}$ evaluates to zero when $p_M  - p_k $.

\end{document}